\documentclass[usenatbib]{mn2e}
\usepackage{color} 
\usepackage{graphics}
\usepackage{epsfig}
\usepackage{natbib}

\begin{document}

\title[Outer disk stellar populations] 
{A study of outer disk stellar populations of face-on star-forming galaxies in SDSS-IV MaNGA:   
causes of H$\alpha$ deficiency}

\author [G.Kauffmann] {Guinevere Kauffmann$^1$\thanks{E-mail: gamk@mpa-garching.mpg.de}\\
$^1$Max-Planck Institut f\"{u}r Astrophysik, 85741 Garching, Germany}

\maketitle

\begin{abstract} 
Integral field unit (IFU) spectra of face-on star-forming galaxies from the MaNGA
survey are stacked in radial bins so as to reach a S/N  high enough to measure
emission lines and Lick indices out to 2.5-3 R$_e$.  Two thirds of galaxies have
stellar populations in the outer disks that are older, more metal poor and less dusty
than in the inner disks.  Recent bursts of star formation have occurred more
frequently in the outer disk, but extinction-corrected H$\alpha$ equivalent
widths are significantly lower at fixed D$_n$(4000) in these regions. I examine
the  properties of a subset of galaxies with the the most H$\alpha$ deficient
outer disks. These regions contain  young stellar populations that must have formed within
the last 0.5 Gyr, but extinction-corrected H$\alpha$ values well below the
values predicted for a standard Kroupa IMF.  The H$\alpha$ deficient galaxies have flat
D$_n$(4000) and H$\delta_A$ profiles with little radial fluctuation, indicating
that star formation has occurred extremely uniformly across the entire disk.
The H$\alpha$ line  profiles indicate that the ionized gas kinematics is also
very regular across the disk. The main clue to the origin of the H$\alpha$
deficiency is that it sets in at the same radius where the dust extinction
abruptly decreases, suggesting a  mode of star formation deficient in massive stars in 
quiescent, HI-dominated gas. Finally, I have carried out a search for galaxies
with signatures of unusual H$\alpha$ kinematics and find that 15\% of the sample
exhibit  evidence for significant ionized gas that is displaced from
the systemic velocity of the disk.
\end {abstract}

\section{Introduction}

The far outskirts of nearby star-forming spiral and irregular galaxies are
of considerable interest because they are hypothesized to be still in the
process of assembly through the infall of gas and accretion of satellite
systems at the present day (Kauffmann 1996; Chiappini et al 1997; Boissier
\& Prantzos 1999; Fu et al 2010). Many  observational studies have focused
on whether the luminosity and/or stellar mass profiles of disk galaxies deviate  
from a pure exponential in the outer disk (Van der Kruit 2001; Pohlen \& Trujillo
2006), and whether radial gradients in mean stellar age and
metallicity also show clear breaks or discontinuities near the edge of the
disk (Bakos et al 2008, Yoachim et al 2012, 
Zheng et al 2015, Ruiz-Lara et al 2016). At least some part of this
phenomenology is currently believed to be result of the redistribution of
evolved stars in the disk through secular processes. Stars that were formed at one
location migrate radially through the disk when they are perturbed from
their orbits by  disk spiral structures or by bar instabilities (Roskar et
al 2008).

The study of young stars and HII regions in outer disks has been another
active avenue of observational research. The launch of the GALEX satellite
enabled a systematic study of so-called extended ultra-violet (XUV) disks
in the local Universe (Gil de Paz et al 2007, Thilker et al 2007). The XUV
terminology first arose because it was noticed that the amount and spatial
extent of star formation in the outskirts of a galaxy can be underestimated
by looking for HII regions alone (as traced by H$\alpha$ emission). In the
prototype XUV galaxy M83, UV-bright stellar complexes are found extending
out to about 4 times the radius at which the majority of HII regions are
detected (Thilker et al 2005).

The physical explanation for the mismatch between the population of young
stars traced by H$\alpha$ and UV observations has been the subject of
considerable controversy. In a classic paper, Martin \& Kennicutt
(2001)  presented an atlas of H$\alpha$ images of 32 nearby spiral galaxies
with well-measured rotation HI and H$_2$ (as traced by CO) surface density
profiles, and had identified prominent breaks in their H$\alpha$ surface
brightness profiles, confirming previous claims of clear star formation
thresholds in disk galaxies. They  also showed that the outer threshold
radii were in general agreement with those expected from the Toomre Q
stability criterion. The discovery of the XUV disk phenomenon led to
considerable speculation that there might be two modes of star formation in
disk galaxies -- a high efficiency mode corresponding to the
gravitationally unstable  inner regions of disks and a low efficiency mode
in the far outskirts (Thilker et al 2007).

The existence of high and low efficiency modes of star formation is now
well established thanks to well-resolved and deeper HI and CO data obtained 
for samples of a few dozen nearby spiral galaxies (Bigiel
et al 2008, 2010). In the inner disks, the bulk of the gas is in the
molecular phase and there is a simple linear relation between star
formation rate surface density and molecular gas surface density. In the
outer disk regions, the molecular fraction drops dramatically and CO is
often undetectable for individual galaxies. Schruba et al (2010) stacked CO
line observations in the outer regions of multiple disk galaxies and found
that the same molecular gas-based star formation law could be extended down
to very low densities, with no evidence for a sharp truncation or threshold
in star formation, as proposed by Martin \& Kennicutt (2010). In this work,
star formation rates were derived using a combination of H$\alpha$, far-UV,
and Infrared (IR) emission at 24 and 70 $\mu$m with standard assumptions
about the form of the stellar initial mass function (IMF). The question of
whether there was any discrepancy between star formation traced by
H$\alpha$ and star formation traced by UV was laid aside.

The so-called H$\alpha$ deficiency problem has, nevertheless, remained the
subject  of ongoing investigation and controversy (see Elmegreen 2017 for a
review). It has been argued that the  decline in H$\alpha$/FUV ratio with
radius reflects a systematic change in the  initial mass function of stars
in disks (Meurer et al 2009, Bruzzese et al 2015).  However, these
inferences have been criticized as being on shaky grounds for a number of reasons:
1) stochastic sampling issues caused by the fact that the most massive stars
are rare compared to less massive stars leading to significant
degeneracies between star formation history IMF (Fumagalli et al 2011), 2) the
fact that a significant component of the H$\alpha$ emission may be in
diffuse form and be missed in some studies (Hunter et al 2011, Lee et al
2011), 3) possible loss of ionizing photons from the galaxy in their very
low density outer regions. 

The most recent work on this topic has made use
of star counts derived from HST imaging of main sequence and red giant
branch stars in the outer disk of M83 to simultaneously constrain the slope
and the upper mass cutoff of the IMF (Bruzzese et al 2020). Averaged over
the 4 HST fields that were available for this galaxy, the main sequence
luminosity function provides some evidence for an IMF deficient in the most 
massive stars.  Together with the H$\alpha$ data, the observations yield
evidence for an IMF with an upper mass cutoff of $\sim$ 20 M$_{\odot}$
compared to values $100-150 M_{\odot}$ in normal star-forming regions. 

In this paper, I analyze outer disk stellar populations for a sample of 126
galaxies with stellar masses in the range $10^{9}-10^{11} M_{\odot}$ with
face-on orientation and with IFU data from the Mapping Nearby Galaxies at
APO (MaNGA) survey that extend out to at least 2.5 $R_e$.  Face-on galaxies
are chosen so as to minimize dust extinction effects.  To ensure high
enough $S/N$ to measure key emission lines and stellar absorption features
in the low surface brightness outer regions of the galaxy, I create stacked
spectra with fixed $S/N$ in increasing bins of $R/R_e$  from the centre of
each galaxy to its outskirts. In comparison to previous work, the analysis of stacked
spectra has the advantage that it does not exclude diffuse emission from lower
surface brightness regions of the disk and that it averages over localized
fluctuations in recent star formation. The analysis also accounts for the
effects of dust by making use of narrow indices such as the 4000 \AA\ break
instead of FUV luminosity. Because the spectra are of high enough $S/N$ to
measure the Balmer decrement, the H$\alpha$ fluxes can also be corrected
for dust extinction. The stacked spectra also have high enough $S/N$ to
measure a  variety of additional properties of the ionized gas and stars,
such as metallicity,  ionization parameter, and emission line shapes that
probe gas kinematics.  Finally, the analysis is carried out for a sample
that is large enough to examine whether there are systematic trends as a
function of galaxy mass, morphological type or global colour.

The paper is organized as follows. Section 2 provides a summary of the data
set and the spectral analysis methodology. Section 3.1 examines how a variety
of spectral properties change between two ranges in radii: $1 < R/R_e < 2$
and  $2< R/R_e <3$. This section examines trends both in individual spectral
properties and in correlations between different properties.  
In Section 3.2, a subsample of
H$\alpha$ deficient outer disks is selected and stellar population
synthesis models are employed to place constraints on both star formation
history and IMF. In Section 3.3, I return to the question of outer disks with evidence
for irregularities in gas kinematics, presenting a number of interesting
cases and discusses prospects for larger samples. Section 4 contains a
summary of the main results in the paper.

\section {The galaxy sample and spectral analysis methods}

The sample of galaxies is drawn from the 15th data release (DR15) of the
Sloan Digital Sky Survey's  MaNGA project (Bundy et al 2015), which is part
of the Sloan Digital Sky Survey IV programme (SDSS-IV; Blanton et al.
2017).  DR15 includes data for 4,621 unique galaxies.  Stellar masses and
structural parameters such as position angles, ellipticities, and
half-light radii for all galaxies in the MaNGA sample are available from
the MaNGA {\em drpall} file; in all cases, the structural parameters are
obtained from the SDSS elliptical Petrosian apertures. All galaxies with
ellipticity parameter $b/a$ greater than 0.75, half-light radii greater
than 2 arcseconds and NUV-r colour less than 4.0 are selected. The NUV-r
colour cut ensures that only galaxies with ongoing star formation are
included in the sample.  The cut on galaxies with near face-on inclination
helps to minimize dust obscuration of young stellar populations in the
outer disks. The cuts yield a sample of 987 galaxies. The SDSS images of
all these galaxies were inspected by eye to find galaxies where the IFU
coverage extends beyond the high surface brightness optical disk, yielding
a sample of 366 galaxies. 
The average half-light radius of the galaxies in
the sample is 3.8 arcsec
and the average redshift is 0.054. The MaNGA fibre diameter on the sky (including
fibre cladding) is 2 arcseconds, corresponding to a physical size of 2.1 kpc.
The number of spaxels interior to the half-light radius  is thus very limited for the galaxies
under study and our  analysis will be confined to spaxels with  $R>R_{50}$.

The available spectroscopic data consists of the raw data from the first
three years of the survey, the intermediate/final data reduction pipeline
(DRP; Law et al 2016) products, and the first release of derived data
products from the data analysis pipeline (DAP;  Westfall et al.  2019;
Belfiore et al. 2019). The spectra cover the wavelength range
360-1000 nm at a resolution  resolution $R \sim 2000$. 
The analysis begins with the model LOGCUBE files,
which provide the binned spectra and the best-fitting model spectrum for
each spectrum that was successfully fit. For DR15, two kinds of LOGCUBE
files have been made available, VOR10-GAU-MILESHC and HYB10-GAUMILESHC.
Only the binning approach differs between the two, and is differentiated by
whether or not the main analysis steps are performed on binned spectra or
individual spaxels. The analysis in this paper uses spectra contained in the
HYB10-GAUMILESHC files, which pertain to individual spaxels. First,  all
the spectra are arranged in increasing order of radial distance from the
centre of the galaxy.  Successive spectra are combined until an average
signal-to-noise per wavelength element of 50 is reached, averaged over the
wavelength range 4500-6000 \AA.  For each galaxy, there are typically
between 10 and 20 stacked spectra in the radial range 1-3 $R_e$

The stellar continuum fitting procedure is documented in detail in
Kauffmann \& Maraston (2019,KM19) and is based on the algorithm described in
Wilkinson et al (2017). The main change made to the procedure described in
KM19 is that instead of fitting combinations of single stellar
populations (SSPs), I have generated a spectral library stellar components
consisting of smoothly varying star formation histories with form ${\rm
SFR}(t) \propto e^{- \gamma t/t_0}$, where $t_0$ is the time when star formation
commences and varies over the range 1-12 Gyr after the Big Bang (for
standard cosmological parameters where the age of the Universe is $13.8
\times 10^9$ yr). $\gamma$ is a time constant that varies from 0 (constant
star formation rate) to -2.  The metallicity of each stellar component
varies over the range from 0.01 solar to solar. Finally, as in KM19,
the two parameter dust extinction model of Wild (2007) is
applied to each component. This has the form: \begin {equation}
\tau_{\lambda}/\tau_v= (1-\mu)(\lambda/5500 \AA)^{-1.3} + \mu(\lambda/5500
\AA)^{-0.7} \end{equation} $\mu$ spans the range from 0 to 1 and $\tau_V$
varies from 0 to 3.

The continuum fitting is only applied to the stacked spectra at radii
greater than 1 $R_e$ from the centre of the galaxy, where the assumption of
a superposition of smooth star formation histories should be a better prior
than a superposition of single stellar populations.  In practice, the
fitting procedure usually converges to very close to its final form in the
first or second iteration. The addition of additional stellar components can
improve the fit, but these components often only contribute at the few
percent level. The most common additional component that needs to be added
is a short duration episode of recent star formation. This will be discussed
in more detail in the next section.

In Figures 1 and 2, I  show example outer disk stacked  spectra over the wavelength
range from 3600 to 7000 \AA. Figure 1 shows a stacked spectrum at $R=2.5 R_{50}$
and Figure 2 shows a spectrum at $R=3 R_{50}$. Cyan shading indicates
1$\sigma$ measurement errors and the red curves show the stellar continuum fits.
The main emission lines are also marked in the figures.

\begin{figure*}
\includegraphics[width=145mm]{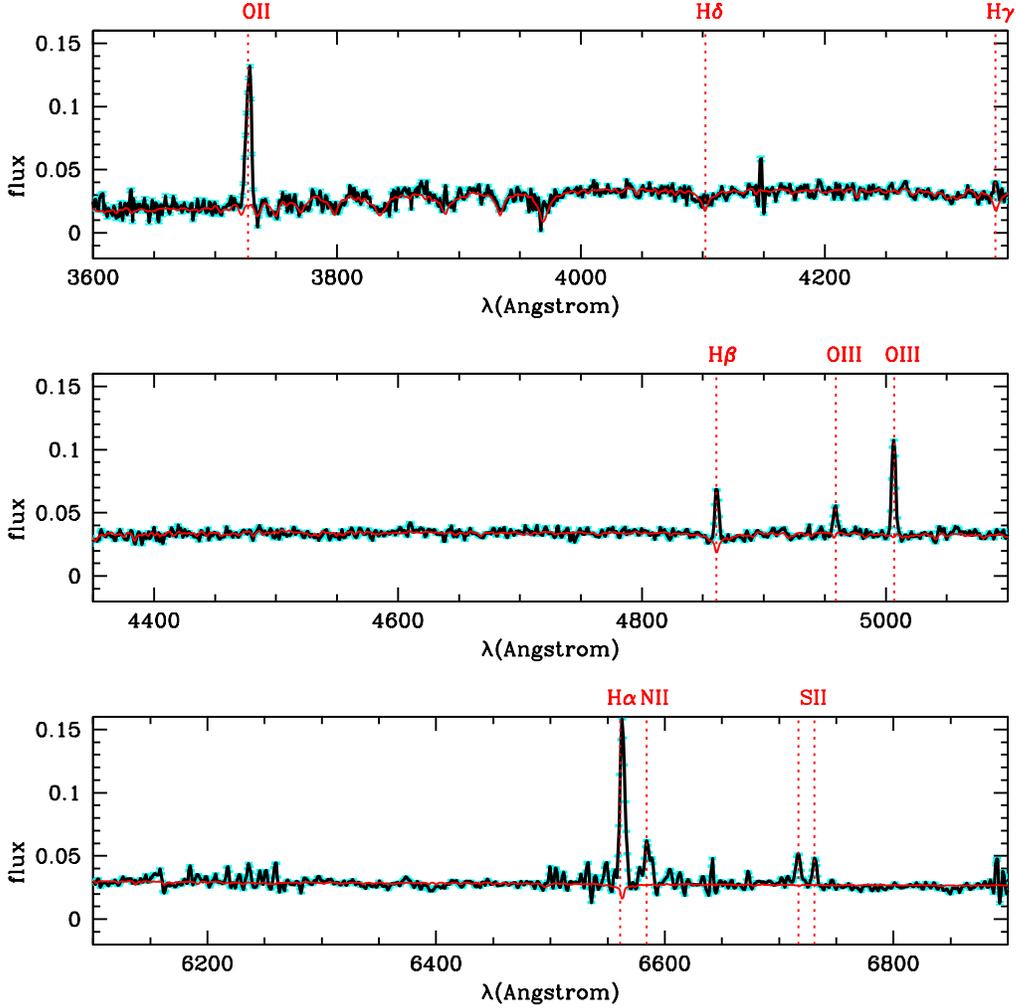}
\caption{An example of an outer disk stacked spectrum at $R=2.5 R_{50}$.
The observed spectrum is plotted in black, cyan shading indicates
1$\sigma$ measurement errors and the red curves show the stellar continuum fits.
The main emission lines are also marked in the figures.
\label{models}}
\end{figure*}

\begin{figure*}
\includegraphics[width=145mm]{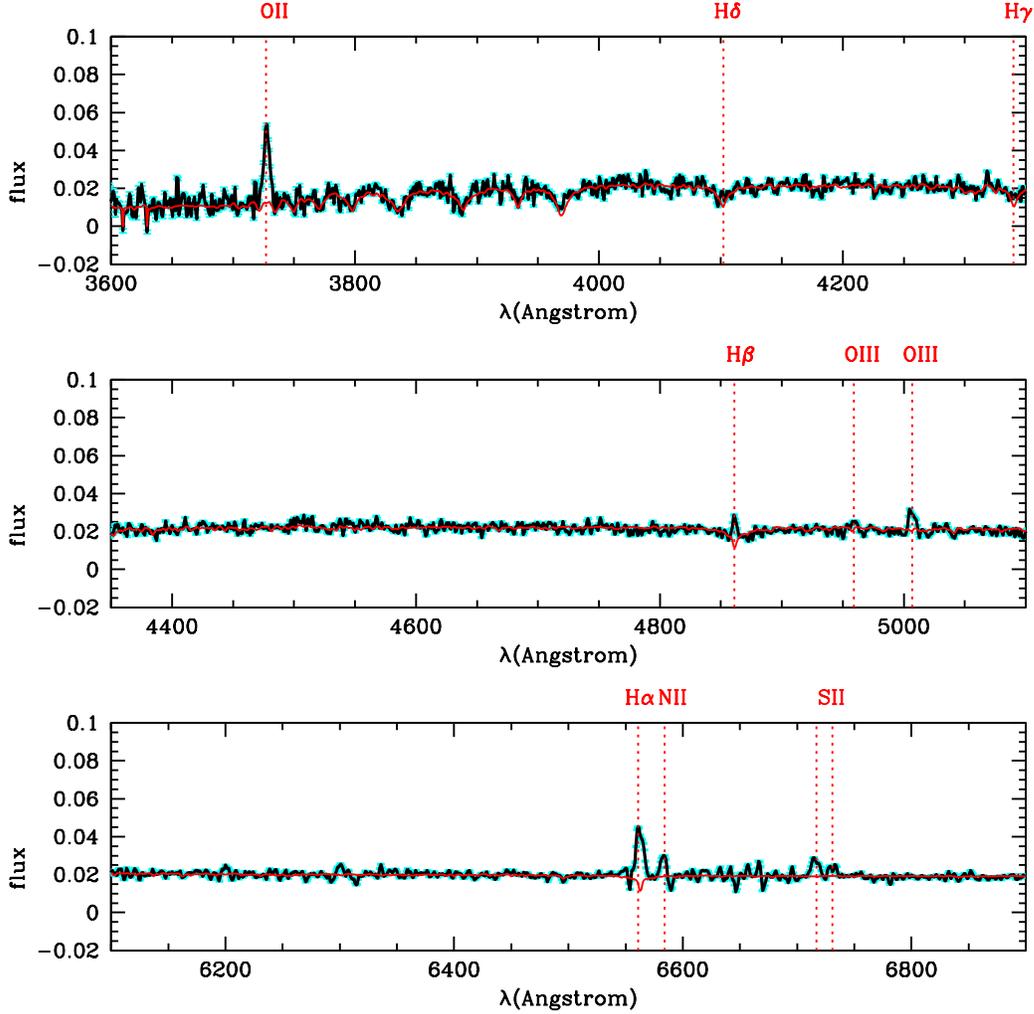}
\caption{As in Figure 1, except for an outer disk stacked spectrum at $R=2.95 R_{50}$.
\label{models}}
\end{figure*}

The best-fit continuum model is then subtracted and single Gaussian line
fitting is carried out for the emission lines [OII]$\lambda$3726,3728,
[NeIII]$\lambda$3869, H$\delta$$\lambda$4102, H$\gamma$$\lambda$4340,
[OIII]$\lambda$4363, [HeII]$\lambda$4686, H$\beta$$\lambda$4861,
[OIII]$\lambda$4959,[OIII]$\lambda$5007, [OI]$\lambda$6300,
H$\alpha$$\lambda$6563, [NII]$\lambda$6583 and [SII]$\lambda$6716,6731

In the far outer disks, the strongest emission line is usually
H$\alpha$$\lambda$6563. Possible deviations from a simple single
Gaussian line shape are characterized by measuring the difference between
the summed flux between 6543 and  6563 \AA\ (after subtracting the best-fit 
stellar continuum model) and half the summed flux predicted by the
single Gaussian line fit. The comparison is made on the bluewards side
of the line centroid so as to avoid any confusion with [NII]$\lambda$6583.   
I define an asymmetry index $Asym$ as 
\begin {equation}Asym= EQW H\alpha- GEWQ H\alpha, \end{equation} where EQW H$\alpha$ is the equivalent
width calculated from the summed flux and GEWQ H$\alpha$ is calculated from
the half-Gaussian.

Finally, the best-fit Gaussian line profiles are subtracted from the
stacked spectra, and the following stellar absorption line indices
are measured: D$_n$(4000), the narrow version  of the 4000 \AA\ break
strength defined in Balogh et al (1999), and the Lick indices H$\delta_A$,
Mgb, Fe5270 and Fe5335. As discussed in Kauffmann et al (2003), the 
location within the plane of H$\delta_A$ versus D$_n$(4000) is a useful
diagnostic of whether recent star formation has occurred in a continuous
manner or in short bursts or episodes. The other three indices form part
of a composite Mg+Fe index, which is sensitive to metallicity (i.e. the 
fraction by mass of all elements heavier than helium over the gas mass)
but shows little sensitivity to $\alpha$/Fe (i.e. the ratio of the total
mass of $\alpha$ elements to the mass of iron). In this paper, 
the index [MgFe]= $\sqrt{{\rm Mgb}(0.72{\rm Fe5270}+0.28{\rm Fe5335})}$ proposed by
Thomas et al (2003) will be used as a stellar metallicity indicator.

\section {Results}

The analysis in this paper follows the same basic route as the one in a
companion paper on the stellar populations in the central regions of nearby
galaxies (Kauffmann 2021, submitted). I first look at the differences
between the stellar populations in the inner disk ($1< R/R_e<2$) and the
outer disk ($2< R/R_e<3$) for the sample as a whole. One of the
main results is that  outer disks
are offset from inner disks in the relation between extinction corrected
H$\alpha$ equivalent width and D$_n$(4000). I then focus on a subclass
of the most H$\alpha$ deficient objects where the relation between H$\alpha$
equivalent width and D$_n$(4000) deviates strongly from the Kroupa (2001)
IMF predictions and look at their properties in more detail.  The main
advantage over previous work is that the analysis in this paper  makes use of results for
stacked spectra, and stellar population properties are thus
smoothed over much larger areas of the galaxy. Note that outer disks
contain much less dust than the central regions of star-forming
bulges, making  uncertainties in fitting stellar population models
to the data less of a concern. 
Only galaxies with 3 or more stacked spectra in the radial range
2-3 $R_e$ are included in the analysis. There are a total of 65 galaxies
with stellar masses in the range $9<\log M_*<10$ and 63 galaxies in the
range $10<\log M_*<11$.

\subsection {Stellar populations in the inner and outer disk} Figure 3
shows differences between a variety of different properties of the
stellar populations and ionized gas in the outer and the inner regions of
the disk for galaxies in the sample with stellar masses in the range
$9<\log M_*<11.5$.  The outer region is defined to be located between 2 and 3
R$_{50}$, where R$_{50}$ is the elliptical aperture half-light radius of
the galaxy listed in the MaNGA {\em drpall}  file, while the inner region
is defined to be located between 1 and 2 R$_{50}$. The difference $\Delta$
is defined by subtracting the inner value from the outer value, so 
that if the outer value is larger, $\Delta$ is positive. In the figure,
$\Delta$ is plotted as a function of the stellar mass of each galaxy.

It is important to understand the extent to which the measured differences
reflect true gradients in physical conditions within the disk, or whether
they are dominated by measurement errors in the line indices and emission
line fluxes and equivalent widths. The errors on the inner and outer disk measurements
are estimated from the variance between the measurements for  individual stacked spectra 
that fall within each radial range.
This takes into account not only the uncertainties due to measurement errors, but
also real variations in physical conditions 
within the disk. The average error on $\Delta$ is indicated in each panel in Figure 3
by a blue error bar. $\Delta$ measurements that are larger than the estimated
error are coloured in magenta.
The two numbers listed in each panel are the number of magenta  points
with positive/negative values, i.e. with higher/lower values of the quantity in
the outer disk compared to the inner disk.

\begin{figure*}
\includegraphics[width=146mm]{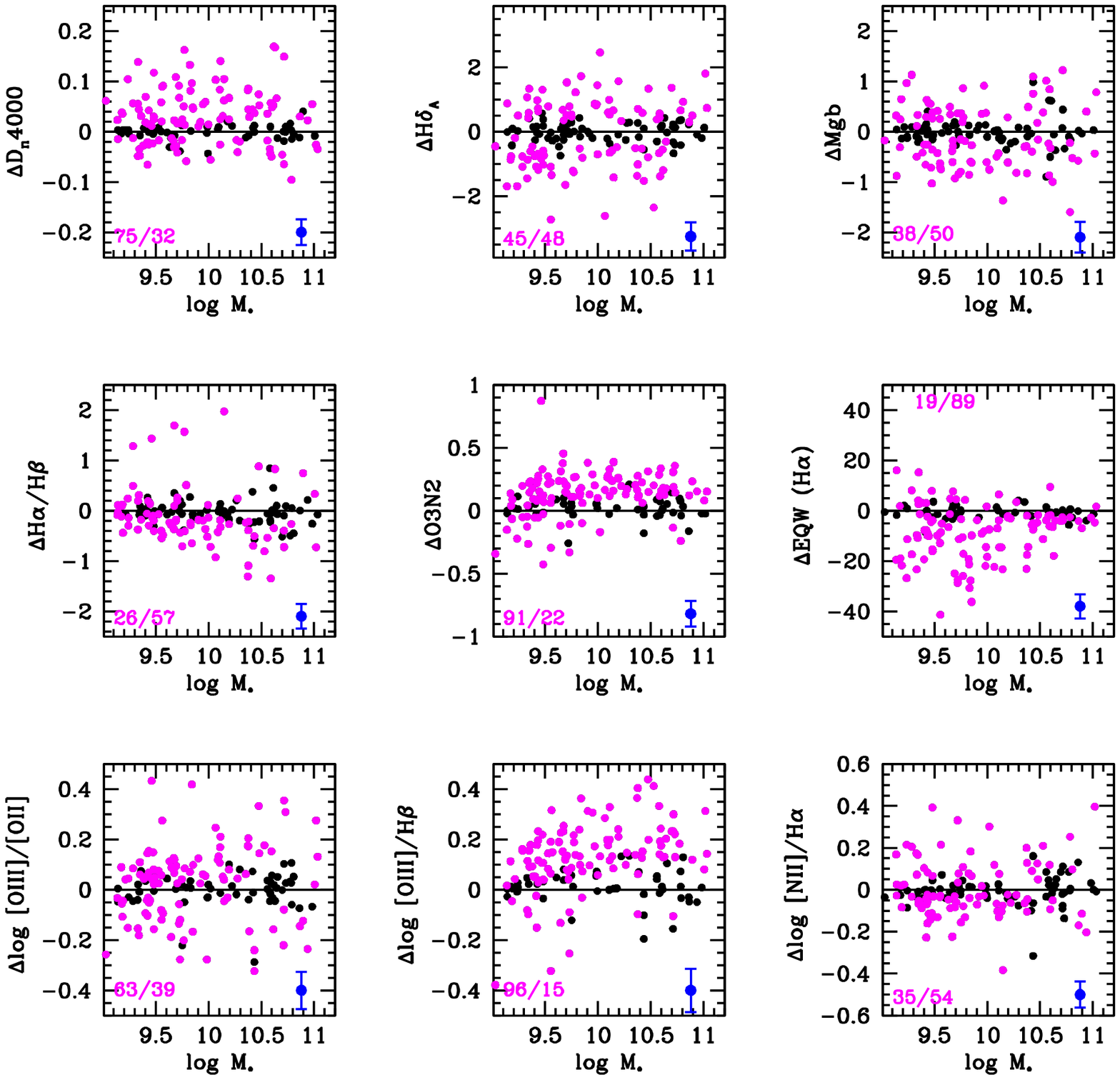}
\caption{ Differences $\Delta$ between a variety of different properties of the
stellar populations and ionized gas in the outer and the inner regions of
the disk sre plotted as a function of stellar mass. 
The outer region is defined to be located between 2 and 3
R$_{50}$, where R$_{50}$ is the elliptical aperture half-light radius of
the galaxy.  The average error on $\Delta$ is indicated in each panel 
by a blue error bar.  $\Delta$ measurements that are larger than the estimated
error are coloured in magenta. The two numbers listed in  in each each panel are the number of magenta  points
with positive/negative values.
\label{models}}
\end{figure*}

\begin{figure*}
\includegraphics[width=145mm]{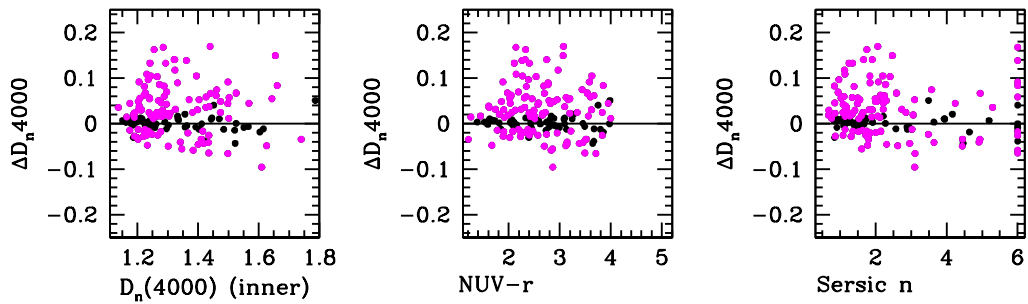}
\caption{As in the previous figure, except that $\Delta$ is plotted as a function
of D$_n$(4000),  Sersic index and NUV-r colour to examine outer disk age gradients as a function
of galaxy structure and disk stellar population properties. 
\label{models}}
\end{figure*}

I now discuss the main findings.
The first panel shows that outer
disks have larger values of D$_n$(4000) than inner disks on average.
This is in accord with previous
studies of radial colour gradients in large sample of galaxies, which find
that the upturn towards redder colours usually occurs at radii greater than
2 $R_e$ (Zheng et al 2015).
An upturn in colour could be caused by either an increase in age, in
metallicity or in dust content at the outskirts of the galaxy. The fact
that Mgb and  H$\alpha$/H$\beta$ are on average lower in the outer disk
(panels 3 and 4) 
demonstrates that stellar metallicity and dust content decrease in the outer disk,
and that the redder outer colours reflect the presence of older stars.
These results are in qualitative
agreement with a study of radial age and metallicity gradients in nearby
galaxies using IFU spectroscopic data from  the CALIFA survey (Gonz\'alez
Delgado et al 2015), which
show that the average age gradient flattens
considerably in the outer disk for later type galaxies in contrast to the
stellar metallicities, which decrease smoothly with radius.

Panel 5 shows the O3N2 emission line index,
introduced by Alloin et al (1979), and defined as \begin {equation} O3N2 =
\log \left(\frac{[OIII]\lambda 5007}{H\beta} \times
\frac{H\alpha}{[NII]\lambda 6583}\right) \end {equation} One popular
calibration (among many) that relates the oxygen abundance and the O3N2
index with a simple linear regression has been proposed by Pettini \& Pagel
(2004)  where $12+\log (O/H)=8.73-0.32\times O3N2$. In this paper I will
only use the  O3N2 indicator as a way of comparing the ionized gas metallicities in a
relative rather than absolute sense. 

The results in panel 5 show that the 
metallicity of the ionized gas is almost always lower in the outer disk than in the inner
disk.
Panel 7 shows the ionization parameter, defined as the ratio of the [OIII]
line luminosity to the [OII] line luminosity. Outer disks have higher
ionization parameter. The results shown in panels 8 and 9 for the 
line rations [OIII]/H$\beta$ and [NII]/H$\alpha$ confirm these findings --
the [OIII]/H$\beta$ rato has the largest outer-inner disk differences because it is sensitive to
both betallicity band ionization.
The radial trends found here for  ionization parameter and gas-phase metallicity agree
well with the analysis of the CALIFA galaxy sample by Rodriguez-Baras et al
(2018).

Panel 6 compares H$\alpha$ equivalent widths in the inner and outer
disks.  No dust correction has been applied. 
H$\alpha$ equivalent widths are
lower in outer disks compared to inner disks for 82\% of the galaxies in
the sample.  This number is roughly consistent with the 70\% of galaxies that
show higher values of D$_n$(4000) in the outer disk (panel 1).  
The H$\alpha$ luminosity can also be corrected for extinction using the formula $A_V=
1.9655 R_V \log(H\alpha/H\beta/2.87)$, where $R_V= 3.1$ is assumed, together
with the Calzetti (2001) extinction curve. 
Applying dust
corrections does not change the  conclusions, it just increases the
difference between the inner and outer disk values.

In Figure 3, I have plotted differences between outer and inner disk quantities
as a function of the stellar mass of each galaxy. The trends are
qualitatively similar across the entire mass range $9 < \log M_* < 11.5$.
More massive galaxies have smaller outer disk  differences in H$\alpha$ equivalent width
and larger differences  in ionization parameter and extinction. 
In Figure 4, I explore how outer-inner differences in the two stellar age parameters
D$_n$(4000) and EQW H$\alpha$ depend on the age of the disk, parametrized
by the value of D$_n$(4000) in the inner disk and the global NUV-r colour
of the galaxy (panels 1 and 2), as well as the Sersic index from a 2 dimensional
fit to the luminosity profile of the galaxy. This figure shows that the bluest
galaxies (D$_n$(4000)$<1.2$ and NUV-r$<2$)  have the smallest differences.     
Galaxies with intermediate morphological type (n=3-5) also have smaller
outer disk differences than very late-type and very early-type galaxies.

I now move on to the correlations between different spectral
quantities and analyze whether they differ systematically between the inner and
outer regions of galaxies. Kauffmann (2021,submitted) showed that the
extinction corrected H$\alpha$ equivalent width in the centres of galaxies
measured at a fixed value of D$_n$(4000) were boosted on average in the
centres of galaxies compared to their disk regions. In a subset of galaxies
where the boost factor was very large, clear Wolf-Rayet signatures were
often found in conjunction with he very high H$\alpha$ equivalent width
values, giving credence to the hypothesis that that an excess of young,  
high mass stars is found in the central regions of  these systems.

Figure 5 shows the relations between H$\delta_A$ and D$_n$(4000)
(panel 1), uncorrected H$\alpha$ equivalent width versus D$_n$(4000) (panel
2), extinction corrected H$\alpha$ equivalent width versus D$_n$(4000)
(panel 3) and extinction corrected H$\alpha$ equivalent width versus
$H\delta_A$ (panel 4).  The red points show individual stacked spectra
measurements from inner disk regions ($1<R/R_e<2$), while blue points show
results from the outer regions ($2<R/R_e<3$). The red and blue lines show the
median of the distributions for red and blue plotted
points.  The median is evaluated by arranging the
measurements in ascending order along the x-axis and evaluating the median
in bins containing a fixed number of galaxies, so that
the up-and-down fluctuations can be read as errors. As can be seen in panel 1,
the median relation between $H\delta_A$ and D$_n$(4000) is the same in the
inner and the outer disk. There are, however, indications that the scatter
in H$\delta_A$ at fixed D$_n$(4000) is larger in the outer disk
compared to the inner disk.  
Because each spectrum in binned to fixed signal-to-noise, the
increase in scatter is likely an indication that  that  star formation histories
in the outer disks  of galaxies have been characterized
by more bursts over the past 1-2 Gyr.

The  best-fit star formation histories
described in section 2 can be used to quantify this in more detail.
Many of the the stacked spectra require more than one exponential to produce
a converged fit. In order to parameterize the star formation histories of
all of the galaxies in a uniform way, I combine the exponentials and calculate
the following quantities:
1) t$_{\rm look back}$(90\%), the look-back time to the epoch before  90\% of the
stars were formed, 2) t$_{\rm look back}$(50\%), the look-back time to the epoch when half the stars were
formed, 3) t$_{\rm look back}$(25\%), the look-back time to when the first quarter 
of the stars were formed.
(3) is a diagnostic of whether there was a recent burst of star formation, (1) is
a diagnostic of the time of the onset of star formation, and (2) is a diagnostic of
the median time of star formation.

Figure 6 shows histograms of these three quantities for the inner disks (red)
and outer disks (blue). The first panel shows that  t$_{\rm look back}$(25\%)
is more frequently a Gyr or less in the outer disk than in the inner disk,
confirming an excess  population of outer disks with recent bursts of star formation. 
The biggest differences between inner and outer disks are seen for the quantity
 t$_{\rm look back}$(50\%). The  median value of this
quantity in the outer disk is at a look-back time of $\sim$ 7 Gyr compared 
to 3.5 Gyr in the inner disk. In contrast, the histograms of 
t$_{\rm look back}$(90\%) are almost identical for the inner and outer disks,
indicating that the look-back time to the onset of star formation is
the same in both environments. The combination of identical distributions
of t$_{\rm look back}$(90\%) and strong differences in t$_{\rm look back}$(50\%)
is in accord with the idea that the physical mechanism causing the
difference is the radial migration of stars in the disk. Older stars have
migrated further than younger stars.

\begin{figure}
\includegraphics[width=92mm]{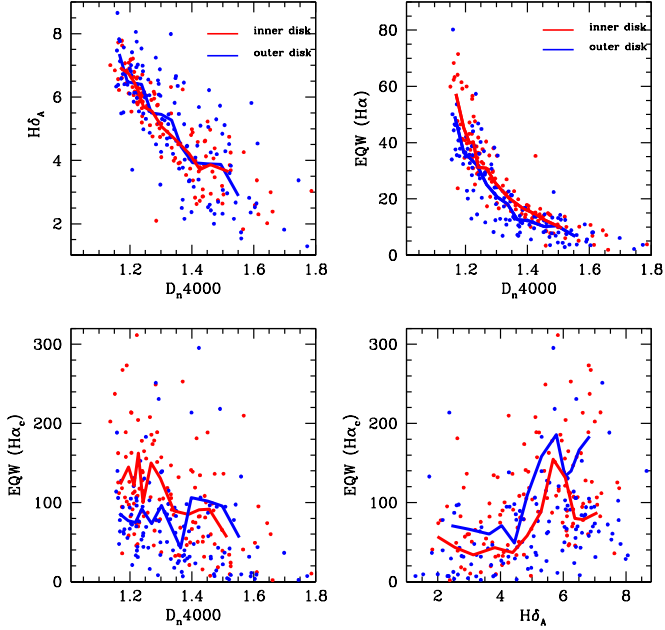}
\caption{The relations between between H$\delta_A$ and D$_n$(4000)
(panel 1), uncorrected H$\alpha$ equivalent width versus D$_n$(4000) (panel
2), extinction corrected H$\alpha$ equivalent width versus D$_n$(4000)
(panel 3) and extinction corrected H$\alpha$ equivalent width versus
$H\delta_A$ (panel 4).  The red points show individual stacked spectra
measurements from inner disk regions ($1<R/R_e<2$), while blue points show
results from the outer regions. Overplotted red and blue lines show running 
medians of the distributions of  the red and blue points. 
The medians are evaluated in bins with a fixed number of points, so
that the up-and-down fluctuations along the x-axis can be read as errors.   
\label{models}}
\end{figure}

\begin{figure*}
\includegraphics[width=145mm]{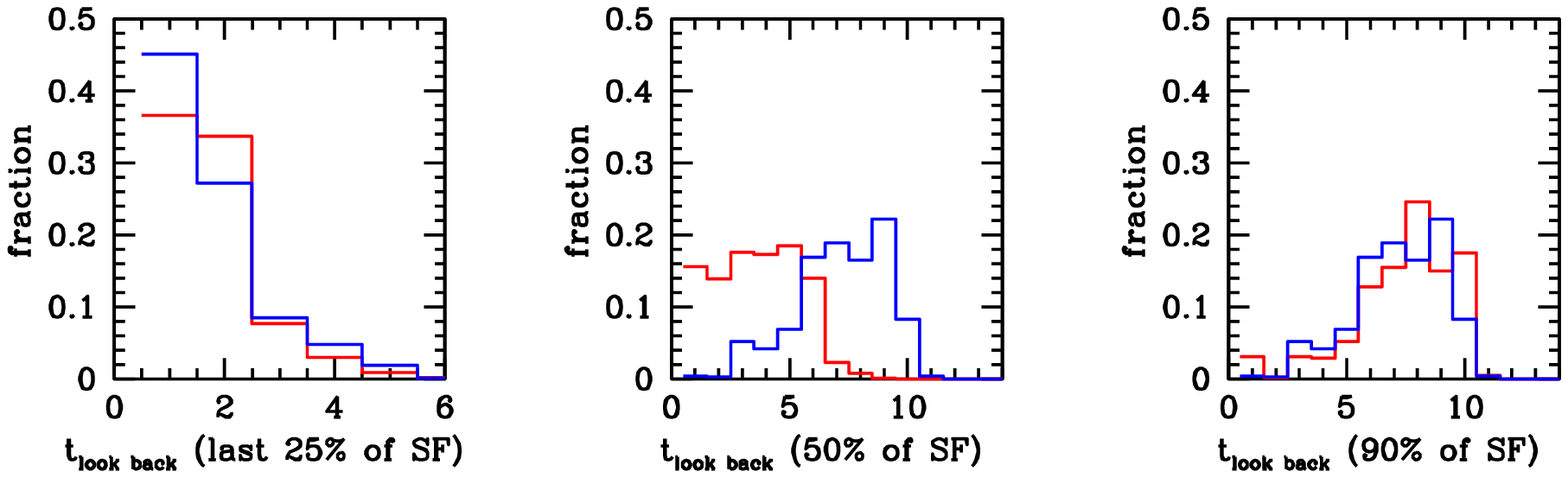}
\caption{Histograms of 1) t$_{\rm look back}$(25\%), the look-back time to when the first quarter
of the stars were formed, 2) t$_{\rm look back}$(50\%), the look-back time 
to the epoch when half the stars were
formed, 3) t$_{\rm look back}$(25\%), the look-back time to when the first quarter
of the stars were formed. Red lines show results for inner disk regions and
blue lines for outer disk regions.  
\label{models}}
\end{figure*}

Returning to the correlations between different indicators of
stellar age, panel 2 of Figure 5 show that   the median
H$\alpha$ equivalent width in the outer disk is displaced to lower values at fixed
D$_n$(4000). This is true for both extinction-corrected and ``raw''
H$\alpha$ equivalent widths, though the effects become larger once dust is
taken into account. The same thing is found for the median relation between
H$\alpha$ equivalent width and H$\delta_A$.  The main conclusion is  that
there is a systematic H$\alpha$ {\em deficiency} in the outer regions of
late-type galaxies. The deficiency is strongest for the galaxies with the
youngest stellar populations (low D$_n$(4000) and high H$\delta_A$).
As will be demonstrated in the next subsection, the D$_n$(4000),
H$\delta_A$ and  H$\alpha$ equivalent widths measured for these 
 outer disk regions  cannot be
fit assuming a standard initial mass function,

Figure 7 shows relations between metallicity indicators and stellar age
indicators, plotted the same way as in Figure 5. The Lick Mgb index and
O3N2 are shown, because these are the
two indicators with the highest $S/N$ measurements in the outer disk.
D$_n$(4000) and extinction-corrected H$\alpha$ equivalent widths are used as
stellar age indicators. The main result is that there is a strong
correlation between age and metallicity in both the inner and the outer
disk, with lower metallicities found at younger ages. The outer disk
exhibits a relation that is slightly offset to lower Mgb and higher O3N2 at fixed
D$_n$(4000) and H$\alpha$ equivalent width.  Note that because star formation
has been more episodic in the outer disks of some galaxies, comparing the relations
at fixed D$_n$(4000) is not equivalent to comparing them at fixed age.
A correction accounting for bursty star formation histories would likely bring the relations into
closer agreement with each other.

\begin{figure}
\includegraphics[width=92mm]{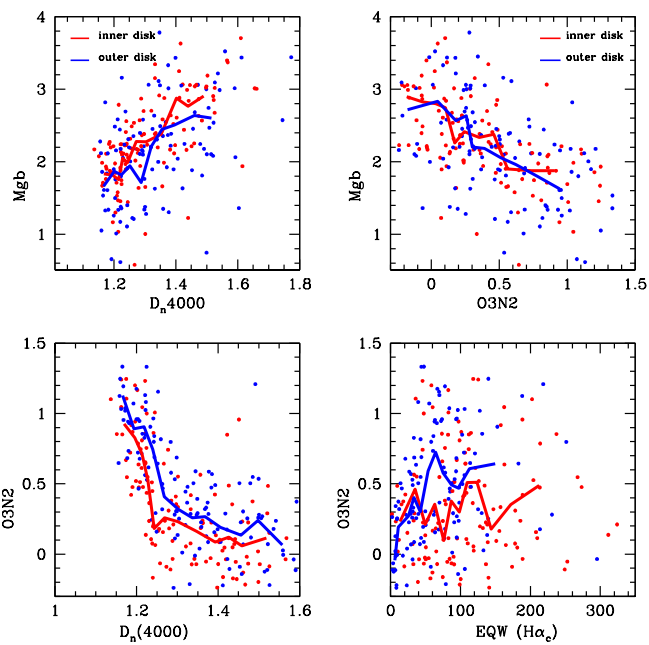}
\caption{The relations between the Mgb Lick index and D$_n$(4000) 
(panel 1), the Mgb Lick index and the O3N2 metallicity indicator for ionized gas (panel
2), O3N2 versus D$_n$(4000) (panel 3), and O3N2 versus                                  
extinction corrected H$\alpha$ equivalent width (panel 4). 
The red points show individual stacked spectra
mesurements from inner disk regions ($1<R/R_e<2$), while blue points show
results from the outer regions. The overplotted red and ble lines show the
medians of the distribution of the red and blue points. 
\label{models}}
\end{figure}

In Figure 8, the top two panels show [OIII]/H$\beta$ versus [NII]/H$\alpha$
BPT line ratio diagnostic diagrams (Baldwin, Phillipps \& Terlevich 1981)
for stacked spectra from the inner (red) and the outer (blue) disks. Results
are shown separately for galaxies in the stellar mass range $9<\log M_*<10$
(left) and $10<\log M_*<11$ (right). There is a strong offset between low
and high mass galaxies, indicative of higher metallicity HII regions in
more massive systems.  There is, however, no clear offset in the locii
occupied by inner disk and outer disk regions, suggesting that the physical
properties of  HII regions are similar across the entire disk.  In the
bottom panels, the Balmer decrement H$\alpha$/H$\beta$ is plotted against
[OIII]/H$\beta$. The Balmer decrement is a measure of the dust extinction
in the HII regions, while [OIII]/H$\beta$ is measure of the ionization
state of the gas. Low-ionization gas in galaxies tends to be dustier.  In
high  mass galaxies, the offset between inner and outer disk regions is
very pronounced, with outer disks containing significantly less dusty,
higher ionization HII regions, than the inner disks.

\begin{figure}
\includegraphics[width=92mm]{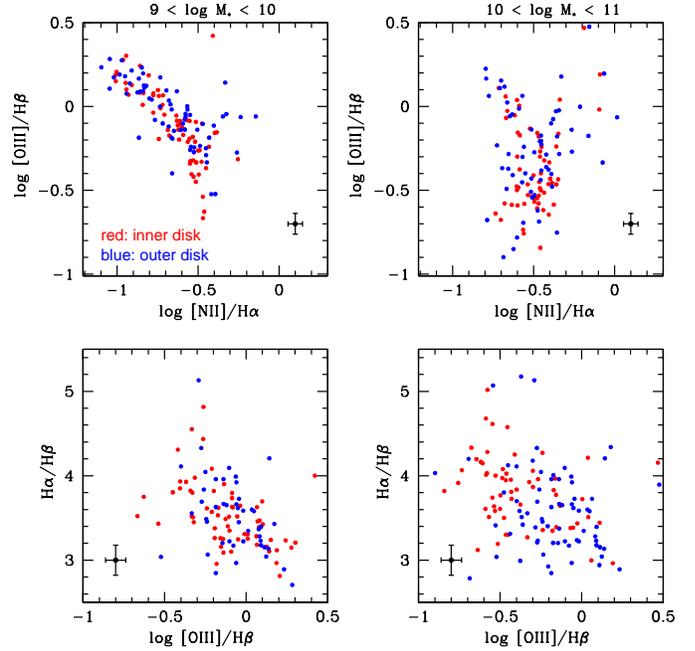}
\caption{The [OIII]/H$\beta$ versus [NII]/H$\alpha$ BPT diagram  
(upper panels) and the Balmer decrement H$\alpha$/H$\beta$
versus [OIII]/H$\beta$ relation .
The red points show individual stacked spectral
mesurements from inner disk regions ($1<R/R_e<2$), while blue points show
results from the outer regions.
Results are plotted separately for galaxies with stellar masses
in the range  $9<\log M_*<10$ (left) and $10<\log M_*<11$ (right).
The error bar in each panel shows the average measurement errors
for the plotted line ratios. 
\label{models}}
\end{figure}

\subsection {Subsample with H$\alpha$ deficient outer disks}

I use the criterion D$_n(4000) < 1.25$ and extinction-corrected H$\alpha$
equivalent width less than 80 \AA\ averaged over the the radial range
$2<\log R/R_e<3$ to select a sample of strongly star-forming galaxies with
H$\alpha$ deficient outer disks. As will be shown later in the section,
stellar populations in this parameter range cannot be fit with a standard
IMF.  There are 21 galaxies (out of a total of 128)  in this subsample.
Figure 9 compares the global properties of this subsample with those of a
``control'' sample of galaxies with  D$_n(4000) < 1.25$ and outer
extinction-corrected H$\alpha$ EQW greater than 80 \AA.  Galaxies are
plotted in the plane of Sersic index versus stellar mass and 
NUV-r colour versus stellar mass.
As can be seen, the galaxies with H$\alpha$ deficient outer
disks span the same range in $M_*$ and in Sersic index as the control
galaxies, showing that they are not unusual in their structural properties.
The H$\alpha$ deficient outer disk sample is slightly shifted to
bluer NUV-r global colours, suggesting that 
the H$\alpha$ deficiency defined using D$_n$(4000) may be the same as
previously studied using XUV disk selection techniques. More detailed
investigation would require analysis of resolved UV phtometry of the
outer regions of these systems.

\begin{figure}
\includegraphics[width=92mm]{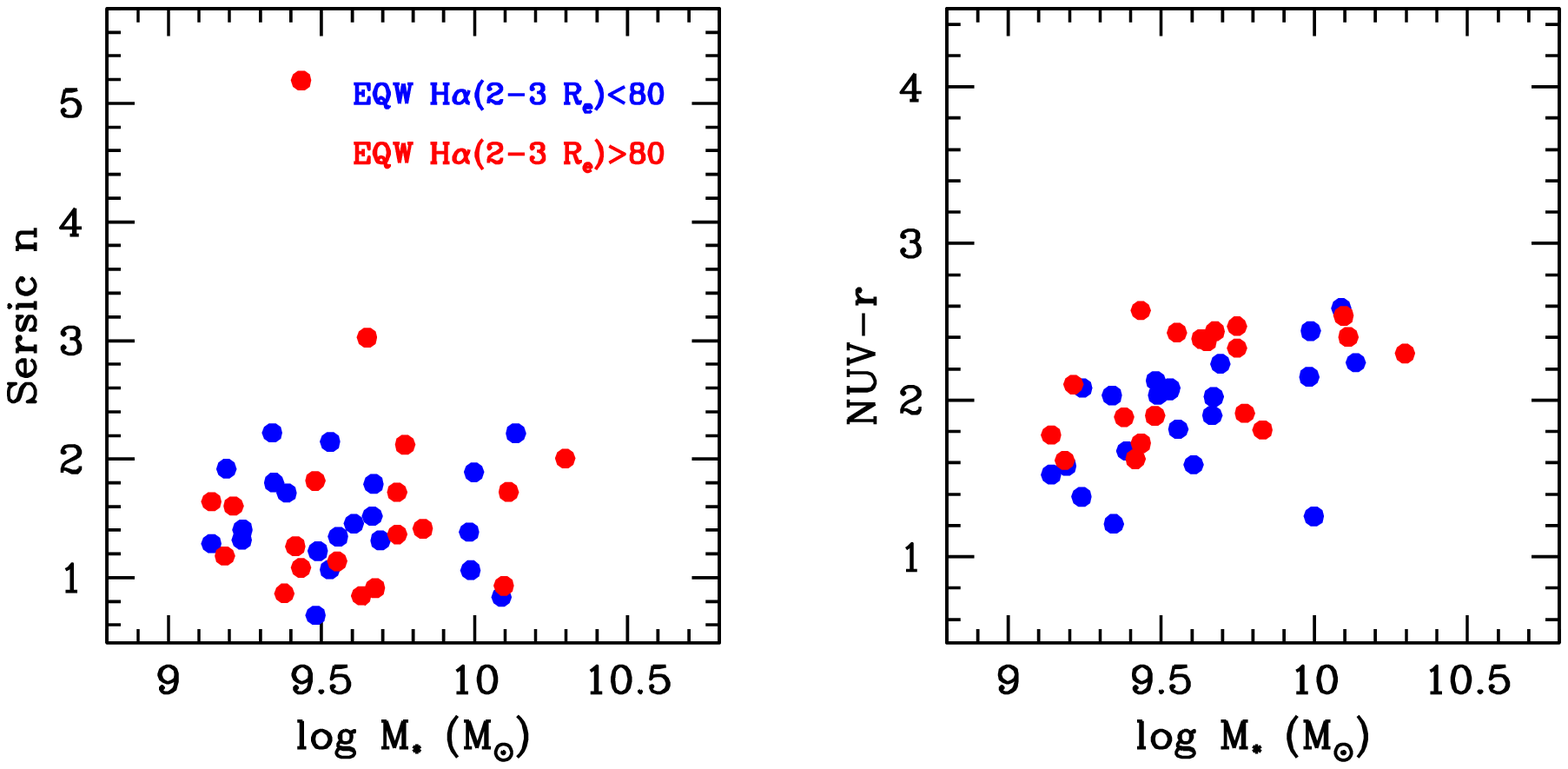}
\caption{{\em Left:} The relation between the Sersic index (from a 1-dimensional fit
to the light profile) and stellar mass is plotted for the H$\alpha$
deficient subsample with D$_n(4000) < 1.25$ and extinction-corrected H$\alpha$
equivalent width less than 80 \AA\ in the outer disk (blue points), 
and for a control sample with D$_n(4000) < 1.25$
and EQW H$\alpha>$80 \AA\ (red points). {\em Right:} The relation between
NUV-r colour and stellar mass for the same two subsamples.   
\label{models}}
\end{figure}

Figure 10  demonstrates that the same basic conclusions reached by Bruzzese
et al (2020) in their analysis of the stellar populations of four HST
fields in the outer XUV disk of M83 also apply to the outer disks of the
galaxies in the subsample selected from MaNGA using optical spectral
diagnostics.  In each of the three panels in the figure, the black points
are individual measurements from stacked spectra in the radial range
$2<R/R_e<3$.  The coloured curves in the  left panel indicate the loci
occupied by continuous models where stars form with a standard Kroupa
(2001) IMF.  As in Kauffmann (2021), the Starburst99 code (Leitherer et
1999; 2014) is used  to predict the H$\alpha$ equivalent width of galaxies
and models from Bruzual \& Charlot (2003) to predict the stellar continuum
indices D$_n$(4000) and H$\delta_A$.  Red, green and blue curves show
results for solar, 0.5 solar and 0.25 solar metallicity models.  In these
models, galaxies begin forming stars 9 Gyr in the past and their star
formation rate is parameterized as $SFR(t)= C e^{(\gamma t)}$, where
$\gamma$ varies over the range -2.0 to 0.0, i.e. from steeply declining to
constant star formation rate as a function of time. As can be seen, none of
these models reach D$_n$(4000) values less than 1.25 and they also
overpredict the H$\alpha$ equivalent width. The cyan curve shows 0.5 solar
continuous models where stars begin to form 0.5 Gyr in the past. These
models do a better job of spanning the observed range in D$_n$(4000), but
still overpredict H$\alpha$.  It should be noted that Bruzzese et al (2020)
found that the main sequence population of the M83 outer disk was well fit
assuming constant star formation over the same 0.5 Gyr timescale, but that
H$\alpha$ was similarly overpredicted.

\begin{figure*}
\includegraphics[width=145mm]{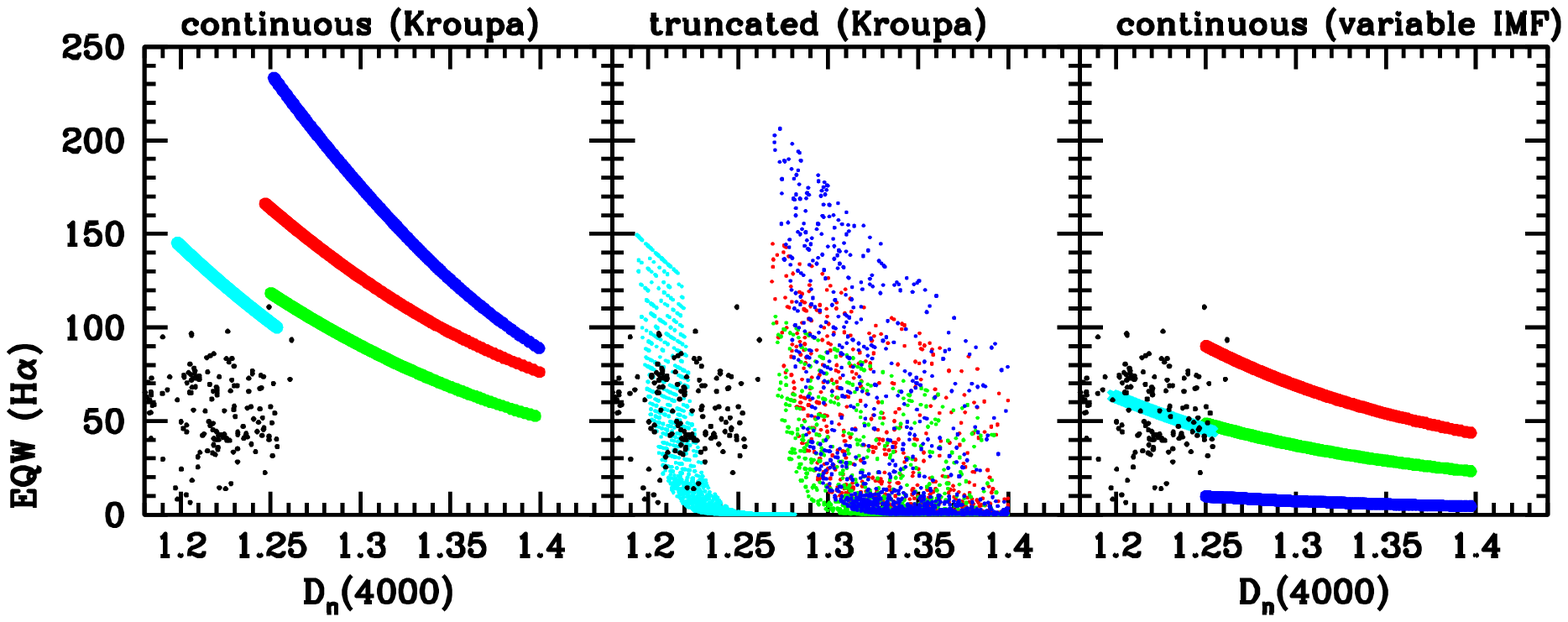}
\caption{{\em Left:} The coloured curves  indicate the loci
occupied by continuous models where stars form with a standard Kroupa
(2001) IMF for 13 Gyr. The cyan curve shows 0.5 solar
continuous models where stars begin to form 0.5 Gyr in the past. 
{\em Middle:} The middle panel shows the 
same set of models where star formation is
suddenly truncated at a random time between 0 and 2$\times10^7$ years in
the past. {\em Right:} A set of continuous models with half
solar metallicity where the Kroupa IMF slope parameters are held fixed, but
where  the upper mass limit of stars is varied (red is for an upper limit
of 50 M$_{\odot}$, green for 35 M$_{\odot}$ and blue for 20 M$_{\odot}$.)
The cyan curve is the model with an upper mass limit of 35 M$_{\odot}$
where stars begin forming 0.5 Gyr in the past.    
In each of the three panels in the figure, the black points
are individual measurements from stacked spectra in the radial range
$2<R/R_e<3$. 
\label{models}}
\end{figure*}

The middle panel shows the same set of models where star formation is
suddenly truncated at a random time between 0 and 2$\times10^7$ years in
the past. Because star formation shut down so recently, residual H$\alpha$
emission from massive stars is still present. As can be seen, these models do
overlap the observed data points. There are two main problems with recent
truncation being a viable scenario: 1) Because the H$\alpha$ emission decays
very quickly, a steadily increasing fraction of galaxies  are found as the
H$\alpha$  equivalent width decreases, whereas the number of galaxies as a
function of H$\alpha$ EQW is roughly constant in our sample, 2) As I show
below, the  H$\alpha$ EQW is roughly constant across the entire outer disk
in our sample of galaxies. Very rapid recent shutdown of star formation
across such a large region violates causality constraints. Bruzzese et al
(2020) similarly argue that recent stochastic fluctuations in star
formation are highly unlikely in M83 because the H$\alpha$ deficiency is
present in all four widely separated HST fields that they study.

The right panel of Figure 10 shows a set of continuous models with half
solar metallicity where the Kroupa IMF slope parameters are held fixed, but
where  the upper mass limit of stars is varied (red is for an upper limit
of 50 M$_{\odot}$, green for 35 M$_{\odot}$ and blue for 20 M$_{\odot}$).
The cyan curve is the model with an upper mass limit of 35 M$_{\odot}$
where stars begin forming 0.5 Gyr in the past.  The main conclusion is that
models where the outer disks begin forming stars recently with an IMF that
lacks stars with masses greater than $\sim$ 35 M$_{\odot}$ can fit the
data.

Figure 11  shows radial profiles for 6 different quantities for these
galaxies.  The radial profiles of D$_n$(4000) of all the galaxies are
extremely flat, varying by at most 0.05 across the entire disk.  The radial
profiles of  extinction-corrected H$\alpha$ width shown in panel 2 separate
into two classes: 1) steeply falling radial profiles with central H$\alpha$
EQW width values of 100 or greater, coloured in red in each panel ; 2) flat
radial profiles, where the H$\alpha$ deficiency persists into the central
regions of the galaxy, coloured in black in each panel.  The galaxies with
steeply falling H$\alpha$ EQW width profiles tend to have stronger dust
extinction (panel 5) and higher gas-phase metallicities (panel 6) in their
central regions. The two classes of galaxies with  H$\alpha$ deficient
outer disks do not appear to have had significantly different star
formation histories over the past 1-2 Gyr as evidenced by the fact that
their H$\delta_A$ profiles are quite similar (panel 3). Finally, panel 4
investigates whether there is any evidence for disturbed gas kinematics in
the H$\alpha$ deficient galaxies. As discussed in section 2, deviations
from a simple Gaussian line profile shape can be probed by comparing the
flux predicted by the best-fit Gaussian with the actual summed flux. The
results plotted in panel 4 show that the line flux differences normalized
by the continuum do not exceed values greater than a few \AA\ even in the
far outer disk. We will show in the next subsection that there are galaxies
in the parent face-on disk sample with outer H$\alpha$ profiles that
deviate much more strongly from a Gaussian, but these galaxies do not have
H$\alpha$ deficient outer disks. Our analysis thus yields no evidence that
the H$\alpha$ deficient outer disk phenomenon is connected with the
accretion of cooler gas traced by H$\alpha$. The fact that only dust
content and metallicity seem to vary between disk regions with and without
H$\alpha$ deficiency suggests instead that molecular gas-phase chemistry
may be the determining factor in this phenomenon.

\begin{figure*}
\includegraphics[width=145mm]{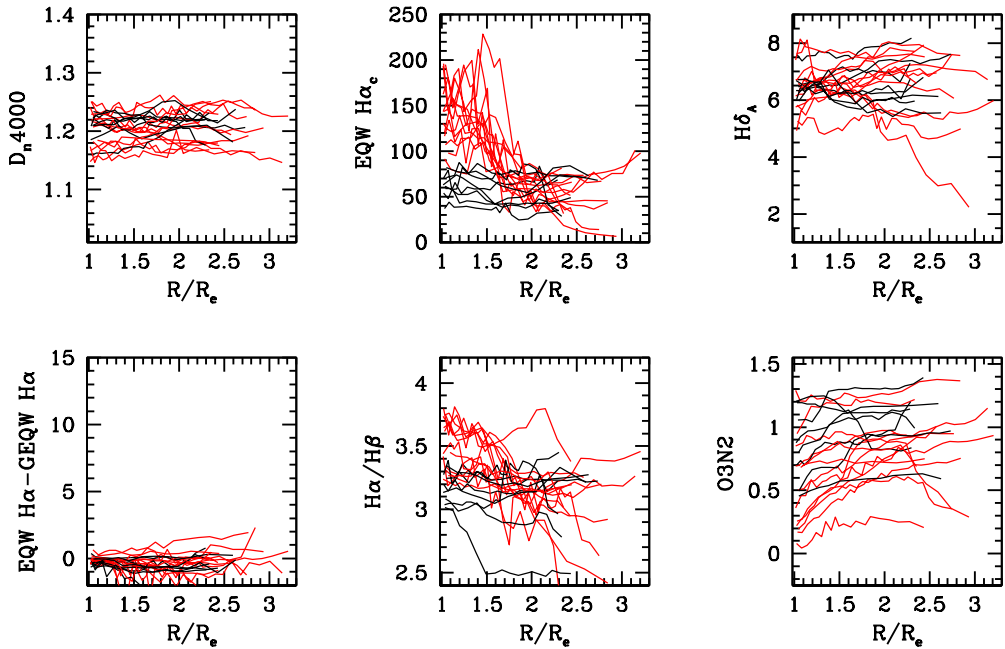}
\caption{Radial profiles of D$_n$(4000), extinction corrected H$\alpha$ equivalent
width, H$\delta_A$, {\em Aym parameter} , Balmer decrement H$\alpha$/H$\beta$
and O3N2 metallicity index for the 21 H$\alpha$ deficient galaxies.
The profiles of galaxies with central H$\alpha$ equivalent widths greater than 100 \AA\
are coloured in red.                                  
\label{models}}
\end{figure*}

\subsection {Subsamples with complex H$\alpha$ kinematics} As detailed  
in section 2, the continuum normalized difference between the H$\alpha$ line
flux given by the best-fit single Gaussian and the flux integrated over the
blue side of the H$\alpha$ spectral window, provides a first-order
diagnostic of kinematic irregularities. This quantity has been calculated
for all the stacked spectra, and the average normalized H$\alpha$ asymmetry
value, which I call {\em Asym} from now on, is evaluated for all spectra
in the radial ranges $1<R/R_e<2$ and $2<R/R_e<3$.  114 out of 133 galaxies
have average {\em Asym} smaller than 3 \AA\ in both the inner and the outer
disk, i.e. 86\% of all galaxes show little or  no evidence for disturbed H$\alpha$
kinematics. Of the remainder, 6 galaxies have disturbed  H$\alpha$
kinematics only in their outer regions.  9 galaxies both in the inner and
in the outer disk, and 4 galaxies only in the inner disk.

Figure 12 shows radial profiles of a variety of quantities for 3 of the
galaxies with the highest {\em Asym} parameters in the  outer disk . As seen in panel 9, in
one galaxy, {\em Asym} rises suddenly to a value of 10 \AA\ beyond 2 R$_e$
and in the other two, it reaches values of $\sim$ 5 \AA\ in the outer
regions.  All three of these galaxies have D$_n$(4000) profiles that
decrease with radius and H$\delta_A$ profiles that increase with radius,
indicating that the outer stellar populations are young. In two of the
galaxies where {\em Asym} rises suddenly at $R=2 R_e$,  H$\delta_A$ also
rises discontinuously to large values at the same radius, but D$_n$(4000)
remains constant. This trend can be understood in terms of an increasing
post-starburst contribution in the outer disk. In all three galaxies, the
stellar metallicity probed by the Mgb and [MgFe] indices decreases with
radius, as does the dust extinction as measured from the Balmer decrement
H$\alpha$/H$\beta$.

\begin{figure*}
\includegraphics[width=145mm]{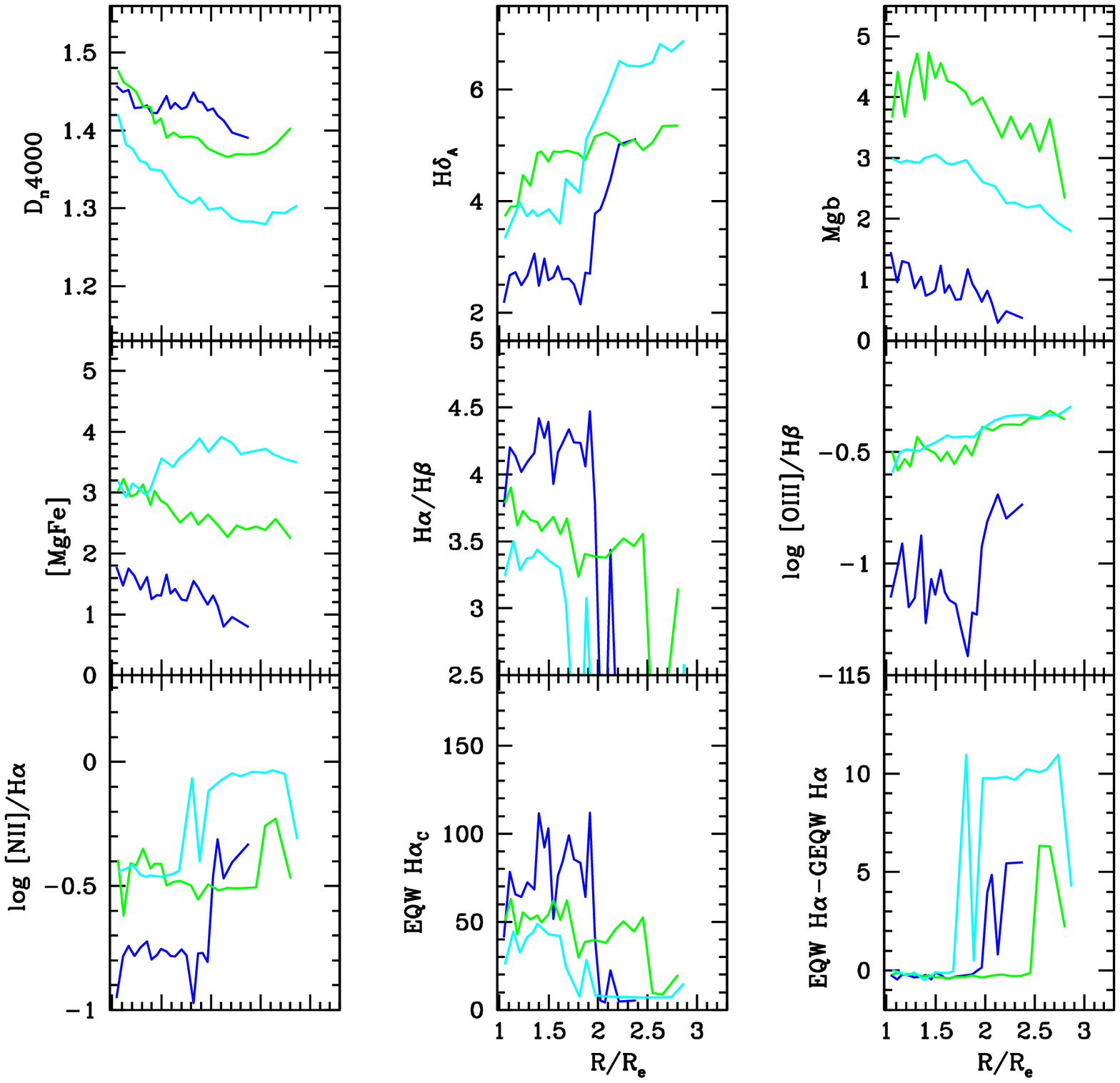}
\caption{Radial profiles of D$_n$(4000), 
H$\delta_A$, Lick Mgb, [MgFe], Balmer decrement H$\alpha$/H$\beta$,
[OIII[/H$\beta$, [NII]/H$\alpha$, extinction-corrected H$\alpha$ equivalent width,
and {\em Asym} parameter for 3 of the galaxies with the highest H$\alpha$ line
asymmteries 
in the outer disk. 
\label{models}}
\end{figure*}

Huang \& Kauffmann (2013) probed the nature of episodic star formation in
nearby galactic disks using a sample of 200 galaxies with longslit spectra.
They found that recent star formation in outer disks is strongly correlated
with the global atomic gas fraction of the galaxy (but not its molecular
gas fraction), and proposed that outer episodic star formation is triggered
by gas accretion events.  The results shown in Figure 6     
corroborate the finding that the stars in the outer disks of nearby
galaxies have often formed in episodic bursts rather than smoothly. I have
checked that selecting a sample galaxies with enhanced H$\delta_A$ at fixed
D$_n$(4000) in the outer disk does not preferentially pick out galaxies
with H$\alpha$ kinematic asymmetries. Evidently, episodic star formation in
outer disks can have a number of different  triggering mechanisms in
addition to gas accretion. Bursts of star formation in the outer regions of
the galaxy could be excited by spiral density waves and other gas
compression mechanisms that are internal, rather than external to the disk.

Figure 13 shows the H$\alpha$ line in the stacked spectra at 6 different
radii for two of the galaxies with high {\em Asym} parameter in the outer
disk.  As can be seen, the line shapes first become flat-topped and then
appear to split into two components. Examination of  SDSS $g,r,i$-band
postage stamp images for  these galaxies reveals
rather regular systems  with no strong asymmetric features and no very
obvious ongoing interactions with a satellite system of significant mass.
Interestingly, the rest-frame H$\alpha$ wavelength is located
exactly in between the two components, which may be an indication of the
presence of a counter-rotating disk rather than a continuous 
distribution of extra planar gas. 
Metallicity variations could also be an interesting diagnostic of the origin of
the extra gas component. 
A metallicity drop was found in one example of an HI-rich galaxy studied
by Moran at al (2010), but
the connection between metallicity, episodic star formation
and kinemetric asymmetries  in outer disks
has not yet been firmly established.
Because of the very small sample size, more detailed analysis of the
kinematic properties of these systems is deferred to future work.

\begin{figure}
\vspace{8mm}
\includegraphics[width=91mm]{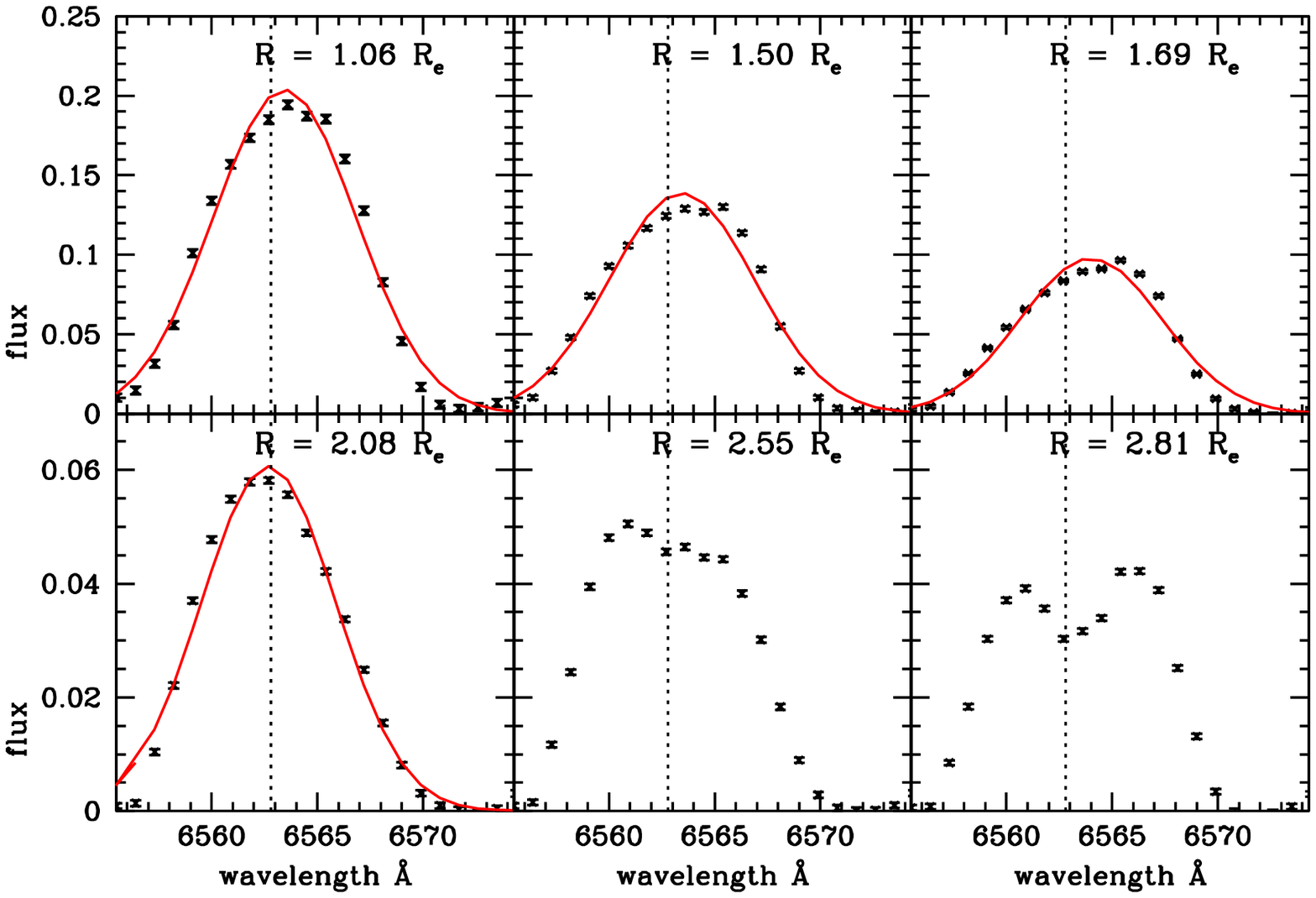}
\includegraphics[width=91mm]{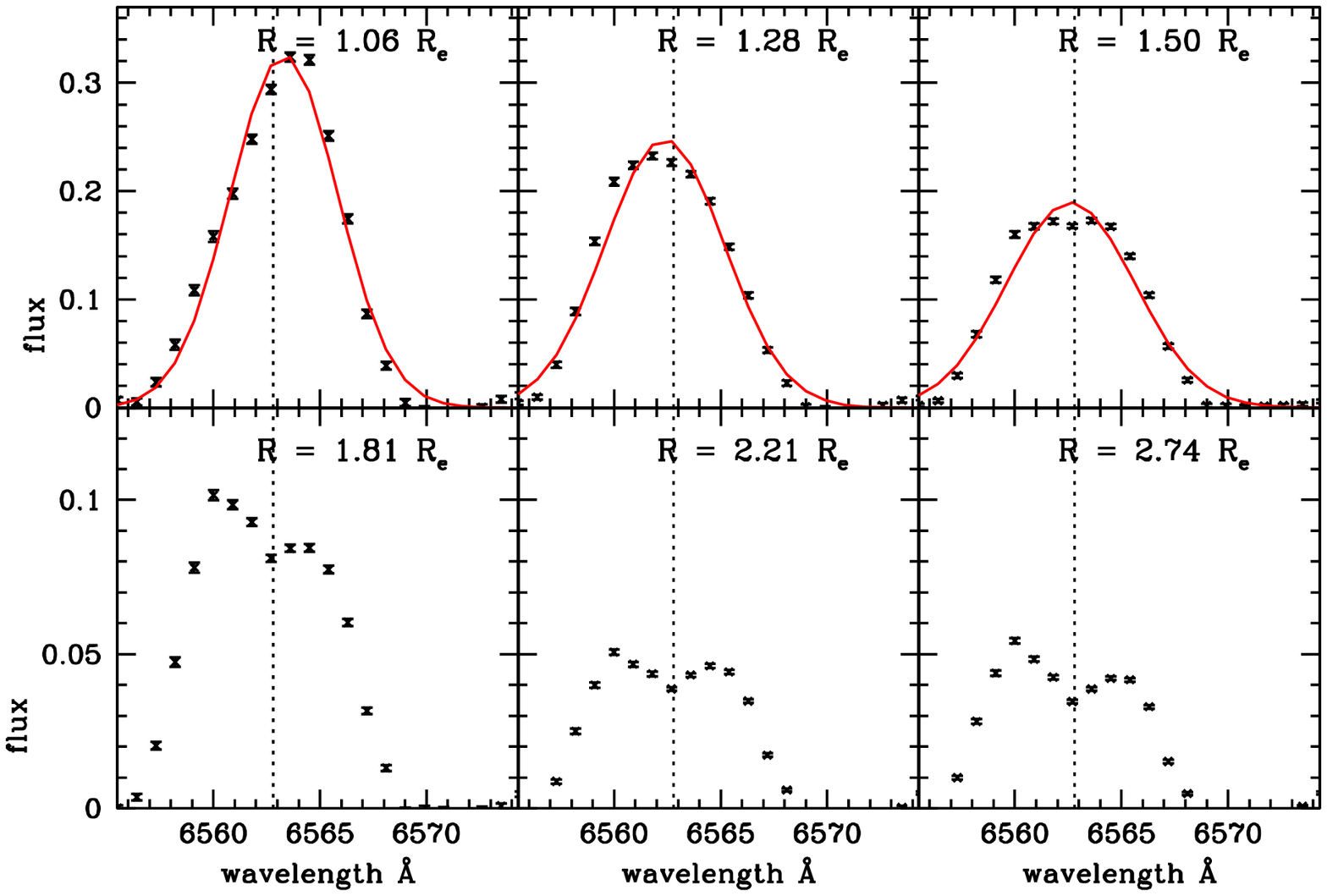}
\vspace{5mm}
\caption{H$\alpha$ line in the stacked spectra at 6 different
radii for two of the galaxies with high {\em Asym} parameter in the outer
disk. 
The dotted line marks the rest-frame  wavelength of H$\alpha$.  
\label{models}}
\end{figure}

I now turn to an examination of the subsample with high {\em Asym} parameter
in both the inner and outer disks. Figure 14 shows radial profiles for the
same quantities as in Figure 12  for three example galaxies. The main difference
with regard to the subsample with high {\em Asym} only in the outer disk
is that there are large fluctuations from one radial bin to another in
the asymmetry parameter, the Balmer decrement H$\alpha$/H$\beta$ and the
extinction corrected H$\alpha$ equivalent width across the entire galaxy.
Stellar ages and metallicities vary more smoothly within each galaxy, 
but there are large differences from one galaxy to another. The H$\alpha$
line profiles plotted for these galaxies in Figure 15 also show large variation in shape from one
radial bin to another. In the inner radial bins, the profiles appear to have 
red or blue wing extensions, while in the outer bins, they appear to be double-peaked,
with the rest-frame H$\alpha$ wavelength located
exactly in between the two components, as in Figure 13.
Finally, I note that the optical images of these systems reveal
no pronounced morphological disturbances or interactions, but the star
forming regions in the innerpart of the galaxy tend to have have considerably 
higher surface brightnesses.

\begin{figure*}
\includegraphics[width=145mm]{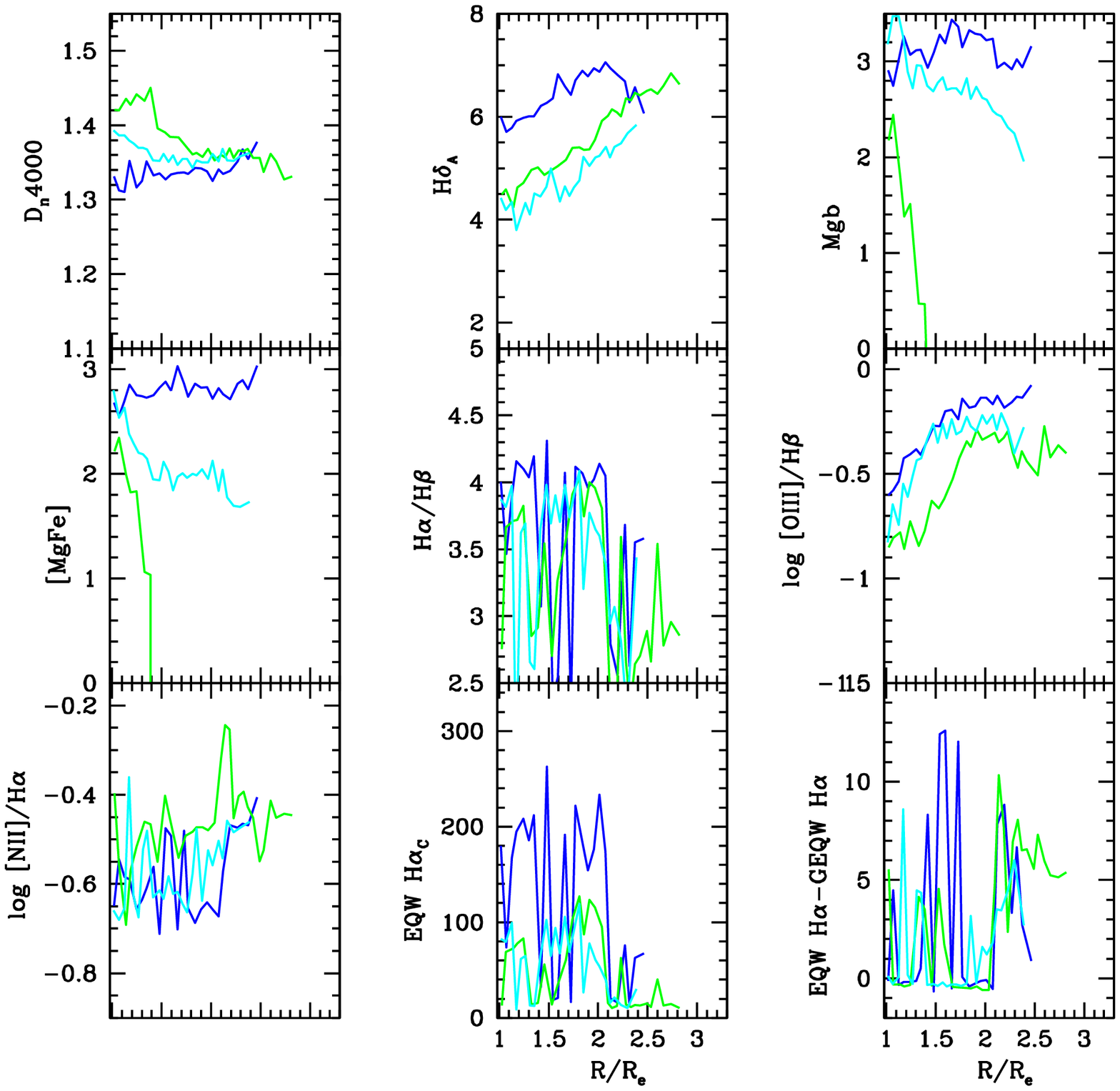}
\caption{Radial profiles of D$_n$(4000), 
H$\delta_A$, Lick Mgb, [MgFe], Balmer decrement H$\alpha$/H$\beta$,
[OIII[/H$\beta$, [NII]/H$\alpha$, extinction-corrected H$\alpha$ equivalent width,
and {\em Asym} parameter for 3 of the galaxies with the highest H$\alpha$ line
asymmteries 
in the inner and  outer disks. 
\label{models}}
\end{figure*}

\begin{figure}
\vspace{8mm}
\includegraphics[width=91mm]{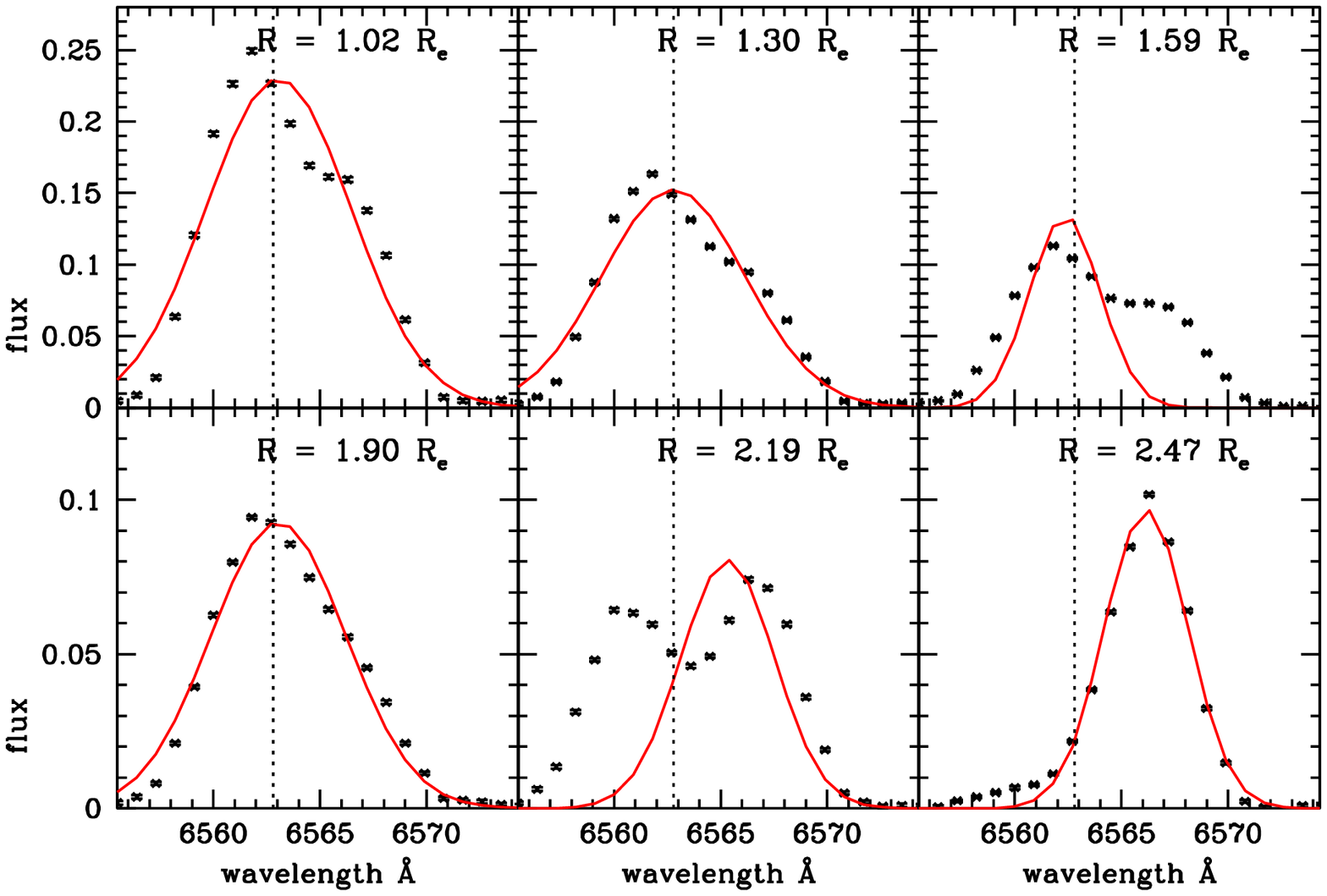}
\includegraphics[width=91mm]{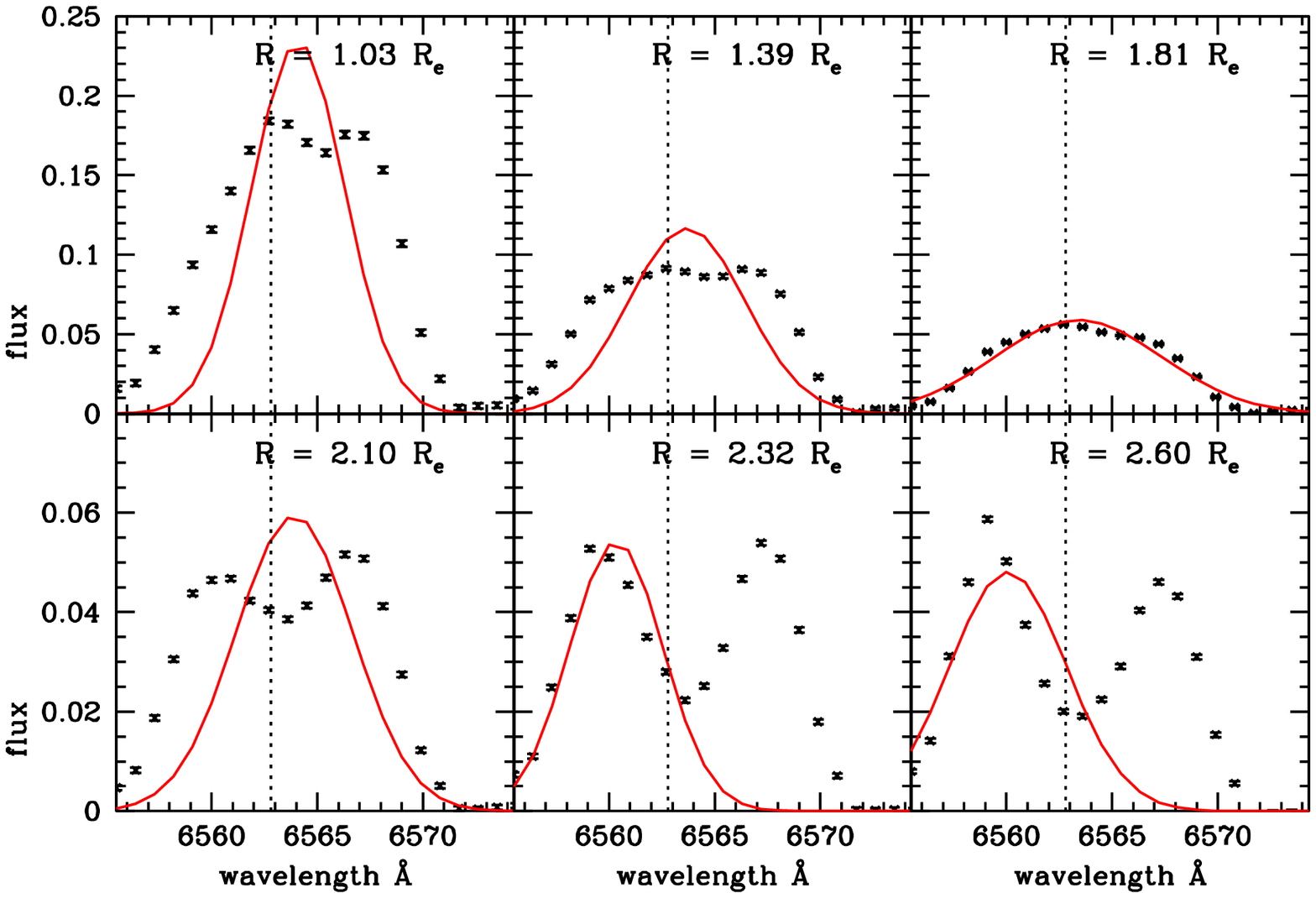}
\vspace{5mm}
\caption{H$\alpha$ line in the stacked spectra at 6 different
radii for two of the galaxies with high {\em Asym} parameter in the inner and outer 
disks. 
The dotted line marks the rest-frame  wavelength of H$\alpha$.  
\label{models}}
\end{figure}

\section {Summary and Discussion}

This paper describes an analysis of the spatially-resolved
stellar populations and
ionized gas properties of a sample of star-forming galaxies with stellar masses in the
range $9<\log M_*<11$.
Integral field unit (IFU) spectra from the MaNGA survey are stacked in radial
bins  to reach a  S/N high enough to measure not only emission lines,
but also the main Lick indices out to 2.5-3 R$_e$.  In this section, I
summarize the main results of the analysis and discuss future directions
for improving understanding of some of the main unresolved issues.
The outer disks differ from the inner disks in the following ways:

\begin{itemize} \item Two thirds of galaxies have outer disk stellar
populations  that are older and more metal poor.  \item Two thirds of
galaxies have lower dust extinction and higher ionization parameters in
their outer disks.  \item 80\% of galaxies have lower metallicities and
H$\alpha$ equivalent widths in their outer disks.  \item Recent bursts of
star formation have occurred more frequently in the outer disk compared to
the inner disk.  \item Extinction-corrected H$\alpha$ equivalent widths are
significantly lower at fixed D$_n$(4000) in the outer disk compared to the
inner disk.  \item Other 2-dimensional correlations between spectral
properties remain roughly constant between the inner and outer disk.  \end
{itemize}

I then examine the  properties of a subset of galaxies with the most
H$\alpha$ deficient outer disks. These are selected to have  $D_n(4000) <
1.25$ in the outer disk, indicative of young stellar populations that must
have formed within the last 0.5 Gyr, but with extinction-corrected
H$\alpha$ values well below the values predicted for a standard Kroupa IMF.
These galaxies all have extremely flat D$_n$(4000) and H$\delta_A$ profiles that
fluctuate very little from one radial bin to the next, indicating that star
formation has occurred extremely uniformly across the entire disk.  The
H$\alpha$ profiles are also well-fit with a single Gaussian at all radii,
indicating that the ionzed gas kinematics is also very regular.  The main
clue to the origin of the H$\alpha$ deficiency is that in a large
fraction of galaxies there is a sharp transition in both H$\alpha$
equivalent width and dust extinction at $R\sim 2R_e$, reminiscent of  the
transition between the inner molecular gas-dominated region of the disk and
the outer HI-dominated region. This suggests that the H$\alpha$ deficiency
may reflect a different mode of star formation in quiescent, HI-dominated gas.

The question of whether stars can form in the absence of molecular gas was
first addressed quantitatively by Glover \& MacLow (2007) and Glover \&
Clark (2012).  These authors ran simulations of the collapse of a gas
clouds to form stars using a set of different chemical models. They showed
that even if molecule formation in the gas was ``switched off'', gas at low
densities is able to cool via C$^+$ fine structure emission. Once at higher
densities, the main cooling process involves the transfer of energy from
gas to dust, so the absence of molecules does not, in and of itself,
constitute a barrier to star formation.  These simulations were highly
idealized experiments and not self-consistent representations of the
chemical enrichment and dust production and destruction processes going on
in the real interstellar medium of a galaxy as stars form and die.  Hu et
al (2016) took the next step of modelling the formation of stars in an
isolated, low-metallicity dwarf galaxy with full treatment of the chemistry
of the ISM, and verified that in such a system, the dense and cold gas
constituting the main reservoir for star formation could be HI rather than
H$_2$-dominated. This simulation did not reach high enough resolution to
probe the formation of stars within individual molecular clouds. The
question of whether a systematic change in the IMF is expected in
star-forming gas that is HI-dominated is still an open one.

I have also carried out a search for galaxies with signatures of
unusual  H$\alpha$ kinematics and find that 15\% of the sample exhibit
clear evidence for significant ionized gas that is displaced away from the
systemic velocity of the main disk. There are a number of detailed studies of
extra-planar HI in face on disk galaxies. Schulman et al (1994)
observed 14 nearly face-on disk galaxies with the Arecibo 305 m telescope
and found the double-horned HI profiles to have high-velocity wings in 10
of these galaxies. Follow-up observations of one of these galaxies, NGC
5668 using the Very Large Array, confirm that 60\% of the material in the
high-velocity wings is distinct in position-velocity diagrams, located
primarily outside the optical disk of the galaxy.  Similar studies have
been carried out for M101 (Van der Hulst \& Sancisi 1988), NGC 608
(Kamphuis \& Briggs 1992), NGC 4254 (Phookun et al 1993), NGC 2403
(Fraternali et al 2002), amongst others.

The current generation of large optical IFU and HI  surveys should enable a
more systematic quantification of the nature of these extra-planar gas
components.  The existing literature is still unclear as to whether the
extra-planar gas is material that is infalling for the first time to fuel
future star formation in the disk, or whether it is ``galactic fountain''
material that was recently ejected by supernovae. Current hydrodynamical
simulations of disk galaxy formation provide detailed predictions for the
way in which present-day disk material assembles. For example, an analysis
of the assembly of 15 Milky Way mass galaxies  from the  Auriga simulations
predicts that half of the stellar material in their present-day disks
originates from gas accreted in subhalo/satellite systems and the other
half is smoothly accreted from the intergalactic medium. About 90\% of all
the material has been ejected and re-accreted in galactic winds at least
once. The vast majority of smoothly accreted gas enters into a galactic
fountain that extends to a median galactocentric distance of 20 kpc with a
median recycling time-scale of 500 Myr (Grand et al  2019). The challenge
for the future is to translate these predictions into observable
diagnostics that can be compared directly with the survey results. This
will be the subject of future work.\\

\vspace{4mm}
{\bf Acknowledgements}\\
\normalsize
Funding for SDSS-IV has been provided by the Alfred
P. Sloan Foundation and Participating Institutions. Ad-
ditional funding towards SDSS-IV has been provided by
the US Department of Energy Onece of Science. SDSS-
IV acknowledges support and resources from the Centre
for High-Performance Computing at the University of
Utah. The SDSS web site is www.sdss.org.
SDSS-IV is managed by the Astrophysical Research
Consortium for the Participating Institutions of the
SDSS Collaboration including the Brazilian Participation
Group, the Carnegie Institution for Science,
Carnegie Mellon University, the Chilean Participation
Group, the French Participation Group, Harvard-
Smithsonian Center for Astrophysics, Instituto de
Astrofsica de Canarias, The Johns Hopkins University
sity, Kavli Institute for the Physics and Mathematics
of the Universe (IPMU)/University of Tokyo, Lawrence
Berkeley National Laboratory, Leibniz Institut fur
Astrophysik Potsdam (AIP), Max-Planck-Institut f\"ur
Astronomie (MPIA Heidelberg), Max-Planck-Institut f\"ur
Astrophysik (MPA Garching), Max-Planck-Institut f\"ur
Extraterrestrische Physik (MPE), National Astronom-
ical Observatory of China, New Mexico State University,
New York University, University of Notre Dame,
Observatario Nacional/MCTI, the Ohio State University,
Pennsylvania State University, Shanghai Astronomical
Observatory, United Kingdom Participation Group,
Universidad Nacional Autonoma de Mexico, University
of Arizona, University of Colorado Boulder, University
of Oxford, University of Portsmouth, University of Utah,
University of Virginia, University of Washington,
University of Wisconsin, Vanderbilt University and Yale
University.

\vspace{4mm}
{\bf Data Availability}\\
\normalsize
Data from this paper will be made available on reasonable request to the
corresponding author.


\end{document}